\documentclass[twocolumn,letterpaper,aps,prc,superscriptaddress,nofootinbib,showpacs,floatfix]{revtex4-2}
\usepackage{color}
\usepackage{xcolor}
\usepackage{indentfirst}
\usepackage{xspace}
\usepackage{verbatim}
\usepackage{epstopdf}
\usepackage[T1]{fontenc}
\usepackage{float}
\usepackage{wrapfig}
\usepackage{graphicx}
\usepackage{amsmath}
\usepackage{amssymb}
\usepackage{amsthm}
\usepackage{array}
\usepackage{tabularx}
\usepackage[utf8]{inputenc}
\graphicspath{ {figures/} }
\usepackage{lineno}
{\large }

\setcitestyle{square, comma, numbers,sort \& compress, super}
\usepackage{appendix}
\usepackage{dcolumn}
\usepackage{bm}
\usepackage{hyperref}
\usepackage[a4paper]{geometry}
\begin{document}
	\title{Event-by-event fluctuations of mean transverse momentum in proton-proton collisions at $\sqrt{s}$ = 13 TeV with PYTHIA8 and HERWIG7 models}
 
	\author{ Subhadeep~Roy }
	\email{subhadeep.roy@cern.ch}
	\affiliation{Indian Institute of Technology Bombay, Mumbai 400076, India}

        \author{ Tanu~Gahlaut }
	\email{tanu.gahlaut@cern.ch}
	\affiliation{Indian Institute of Technology Bombay, Mumbai 400076, India}

	\author{ Sadhana~Dash }
	\email{sadhana@phy.iitb.ac.in}
	\affiliation{Indian Institute of Technology Bombay, Mumbai 400076, India}

	\begin{abstract}

        Estimations of event-by-event mean transverse momentum ($\langle p_{\rm T} \rangle$) fluctuations are reported in terms of the integral correlator, $\langle \Delta p_{\rm T} \Delta p_{\rm T}\rangle$, and the skewness of event-wise $\langle p_{\rm T} \rangle$ distribution in proton$-$proton (pp) collisions at $\sqrt{s}=13$ TeV with the Monte Carlo event generators PYTHIA8 and HERWIG7. The final-state charged particles with transverse momentum ($p_{\rm T}$) and pseudorapidity ($\eta$) ranges $0.15 \leq p_{\rm T}\leq 2.0$ GeV/$c$ and $|\eta| \leq 0.8$ were considered for the investigation. The correlator, $\langle \Delta p_{\rm T} \Delta p_{\rm T}\rangle$, is observed to follow distinct decreasing trends with average charged particle multiplicity ($\langle N_{\rm ch} \rangle$) for the models. Furthermore, both models yield positive finite skewness in low-multiplicity events. Fluctuations are additionally studied using the transverse spherocity estimator ($S_{\rm 0}$) to understand the relative contributions of hard scattering (jets) and other soft processes to the observed fluctuations. Comparing model predictions for $\langle p_{\rm T} \rangle$ fluctuations provides valuable insight into the sensitivity of these fluctuations to hadronization and parton shower models. This is essential for a reliable interpretation of the fluctuation dynamics in pp collisions. Moreover, such comparisons would help to establish a crucial baseline for identifying and studying non-trivial fluctuations in heavy-ion collisions.
	\end{abstract}

	\maketitle

	\section{Introduction} \label{intro}
    The study of event-by-event dynamical fluctuations in quantities like mean transverse momentum ($ \langle p_{\rm T} \rangle$) is an essential tool to investigate the possible formation of QGP in heavy-ion collisions \cite{star2005, alice2014, trainor1}. One of the proposed signs of the QGP formation is a non-monotonic behavior of such fluctuations as a function of centrality or incident energy \cite{heiselberg2001event}. The event-wise $\langle p_{\rm T} \rangle$  can be considered as a proxy for local temperature and hence the study of its event-by-event fluctuation can be associated with the temperature fluctuations in the initial stages, providing relevant information on the phase transition in QCD \cite{stodolsky}. However, fluctuations arising in the initial-state of the colliding nuclei also play a major role, a hint of which was obtained from the measurement by the ALICE Collaboration for Pb$-$Pb collisions at center-of-mass energy per nucleon $\sqrt{s_{\rm NN}}=2.76$ TeV \cite{alice2014}. Recent  measurements by the STAR \cite{star2005,star2019} and ALICE \cite{alice2014} experiment at  RHIC and LHC reported that the strength of dynamical $ \langle p_{\rm T} \rangle$ fluctuations decreased monotonically from peripheral to central collisions in heavy-ion collisions. This observation was attributed to a number of mechanisms, such as the onset of thermalization, collectivity, jet suppression, string percolation, and initial-state density fluctuations \cite{mptvoloshin, schenke}. Furthermore, a recent study of the ATLAS experiment introduced a novel technique to disentangle the effects of the initial-state geometry from the observed fluctuations \cite{atlas2024disentangling}. In pp collisions, the observed fluctuations are expected to be predominantly driven by jets, hadronic resonance decays, and contributions from the underlying events. Although further investigations are encouraged for heavy-ion collision systems, studying $ \langle p_{\rm T} \rangle$ fluctuations in small systems (pp) remains essential for understanding the fluctuation dynamics.
	\par
	In this work, we present a simulation study of $ \langle p_{\rm T} \rangle$ fluctuations in pp collisions at collision energy $\sqrt{s}$ = 13 TeV using Monte Carlo models PYTHIA8 and HERWIG7. Fluctuations in $\langle p_{\rm T} \rangle$ are typically characterized by the second moment of its distribution, providing a measure of momentum correlation by quantifying the deviation of a particle's $\langle p_{\rm T} \rangle$ from the event-wise $\langle p_{\rm T} \rangle$. In addition, investigation of the higher-order moments of the $\langle p_{\rm T}  \rangle$ distribution provides valuable insight and a deeper understanding of the intricate mechanisms underlying the observed fluctuations. Further, the fluctuation observables are studied as a function of transverse spherocity, a quantity known to have the ability to differentiate events based on the dominant mode (hard/soft) of particle production. Estimates for two different Monte Carlo event generators - PYTHIA8 and HERWIG7 were compared to understand the effects of different hadronization mechanisms and other underlying microscopic processes on the observed fluctuations.\par

    The paper is structured as follows. Sec. \ref{model_descriptions} presents a brief description and highlights the key differences between the model PYTHIA8 and HERWIG7. In Sec. \ref{spherocity}, the  transverse spherocity observable is introduced as an event-shape observable. Sec. \ref{observable_definitions} discusses about the observables utilised for this investigation. In Sec. \ref{result_discussions}, the findings of the study are discussed followed by the summary in  Sec. \ref{summary}. 
    
    
    \section{Model Description} \label{model_descriptions}
    PYTHIA8 \cite{pythia8,10.21468/SciPostPhysCodeb.8, 10.21468/SciPostPhysCodeb.8-r8.3} and HERWIG7 \cite{bahr2008herwig++,bellm2016herwig} are two extensively used Monte Carlo (MC) event generators to understand the experimental measurements at the LHC proton-proton (pp) collider. The models implement different mechanisms of hadron formation in the final states. HERWIG7 is based on the \textit{Cluster model} \cite{geiger1993particle} where the colour-singlet $(q\bar{q})$ combinations lead to the formation of clusters which subsequently decay into final pairs of hadrons \cite{field1983qcd}. In contrast, PYTHIA8 is based on the \textit{String fragmentation model} \cite{andersson1983general,andersson1980semi} . The model is built on the dynamics of relativistic strings that produce a linear confinement potential between $q\bar{q}$ pair. As the color flux stretches between the initial $q\bar{q}$ pair, the string fragments into hadrons. The microscopic process of color reconnection (CR) has also been included in both the models. The inclusion of CR plays a crucial role in obtaining an accurate description of growing $\langle p_{\rm T}\rangle$ as a function of average multiplicity ($\langle N_{\rm ch} \rangle$) \cite{color0, color1, color2}. In addition to the initial hard scattering of the partons, the models incorporate additional softer underlying event (UE) activities like multiparton interactions (MPI) and beam remnants etc. The present investigation uses PYTHIA8 with Monash 2013 tune \cite{skands2014tuning} and HERWIG7 with default minimum bias tune \cite{gieseke2012colour,bartalini2011multi}. 
    Comparing the model predictions of the $\langle p_{\rm T} \rangle$ fluctuations can help constrain the model parameters and improve our understanding of fluctuation dynamics in pp collisions.
    

    \section{Transverse Spherocity} \label{spherocity}
    \begin{figure}[!htp]
		\centering
		\includegraphics[width=0.8\linewidth]{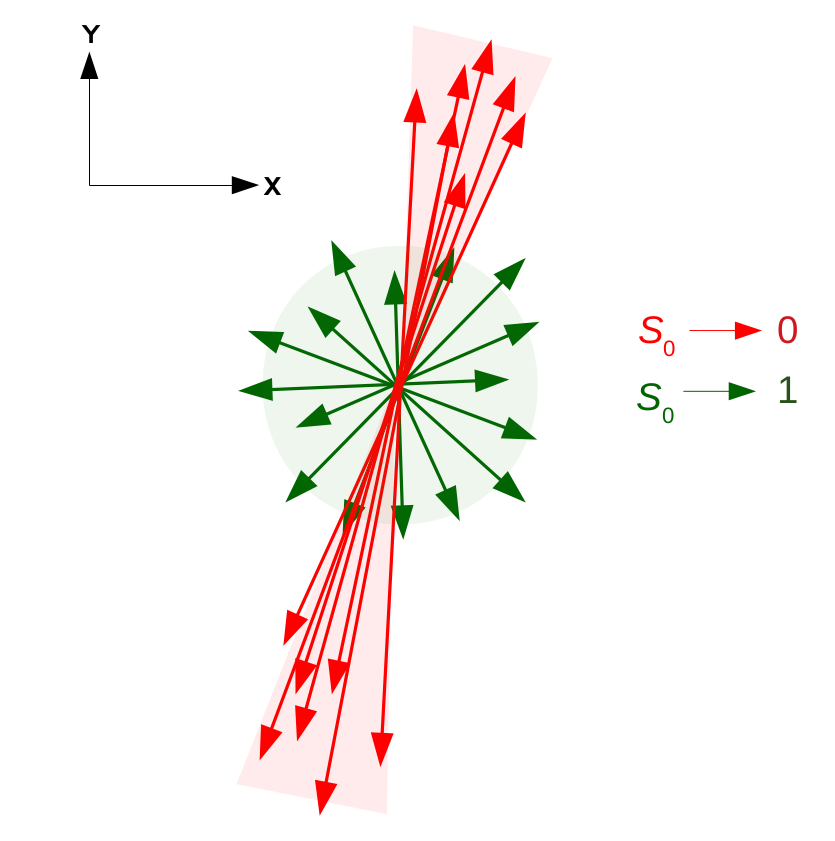}
		\caption{Illustration of the typical topologies for the two $S_{0}$ limits in the azimuthal plane.}
		\label{spherocity_dia}
    \end{figure}
	Event-shape observable like transverse spherocity helps to extract information on the underlying processes of particle production and has been used extensively by various LHC experiments \cite{sphero3, sphero4}. The observable distinguishes events into hard jet-like events (high- $p_{\rm T}$ jets) and soft isotropic events (low $Q^2$ partonic scatterings) based on the distribution of particle's $p_{\rm T}$ on the azimuthal plane \cite{sphero1,sphero2}. The quantity is defined with a unit vector ${\hat{n}(n_{\rm T},0)}$, which minimizes the ratio -
    \begin{equation}
	S_{0}^{|\vec{p}_{\rm T}|= 1} = \frac{\pi^{2}}{4}\underset{\hat{ n}}{\rm min} \left ({\frac{\sum\limits_{i} |\hat{p}_{\rm T_{i}} \times \hat{ n}|}{N_{\rm ch}}}\right )^{2} 
    \end{equation} 

    The condition $|\vec{p}_{\rm T}|$ = 1 removes the neutral jet bias of the observable \cite{alice2024light}. By construction, $S_{\rm 0}$  has limiting values 0 to 1, corresponding to jetty and isotropic events, respectively. Fig. \ref{spherocity_dia} presents an illustration for the two limiting values in the azimuthal plane. Events with a minimum of 5 charged particles with $p_{\rm T}\geq 0.15$ GeV/$c$ and $|\eta|\leq 0.8$ were considered to construct a meaningful geometry using $S_{\rm 0}$. The jetty and isotropic events, representing the extreme 20\% of the spherocity distribution, are selected based on specific $S_{\rm 0}$ limits, as outlined in Table \ref{spherocity_limits}.
 
    \begin{table}[!htp]
    \begin{center}
    \caption{Limiting values for the selections of jetty and istotropic events.}
    \label{spherocity_limits}
    \begin{tabular}{|c|c|}
    	\hline
            Event Class & Spherocity limit  \\
            \hline
    	Jetty (0-20\%) & 0 $\leq S_0 \leq$ 0.595  \\
    	\hline
    	Isotropic (80-100\%) &  0.825 $\leq S_0 \leq$ 1\\
    	\hline
    \end{tabular}
    \end{center}
    
    \end{table}
    \section{Observables} \label{observable_definitions}
    The mean value of the transverse momentum, $\left\langle p_{\rm T}\right\rangle$, calculated in event-by-event basis is expressed as	
    \begin{equation}\label{eq:mean_pt}
	\left\langle p_{\rm T}\right\rangle= \frac{\sum_{i=1}^{N_{\rm ch}} p_{{\rm Ti}}}{N_{\rm ch}} 
	\end{equation}		
    where $N_{\rm ch}$ denotes the total number of charged particles in the given event and $p_{{\rm Ti}}$ represents the transverse momentum of the $i^{\rm th}$ particle in the event. Although, the standard moment method to investigate the $p_{\rm T}$ fluctuations in even-by-event basis provides a comprehensive measure of the total fluctuations accounting for both statistical and dynamical fluctuations, we have used the multiparticle $p_{\rm T}$-correlators to study the event-by-event fluctuations of $\left\langle p_{\rm T}\right\rangle$. The benefit of using multiparticle $p_{\rm T}$-correlators is that they isolate the non-statistical fluctuations of interest by producing zero values for events involving randomly selected particles \cite{Voloshin:2002ku}.
    The two-particle transverse momentum correlator related to the variance of the $\left\langle p_{\rm T}\right\rangle$ distribution is constructed as 
        \begin{widetext}

	\begin{equation}
	\left\langle \Delta p_{\rm T i}\Delta p_{\rm T j}\right\rangle=\left\langle\frac{\sum_{i, j \ne i}^{N_{\rm ch}}(p_{\rm T i}-\left\langle\! \left\langle p_{\rm T}\right\rangle\!\right\rangle ) (p_{\rm T j}-\left\langle\!\left\langle p_{\rm T}\right\rangle\!\right\rangle )}{N_{\rm ch}(N_{\rm ch}-1)} \right\rangle_{\rm ev}	
	\end{equation}

    The average, denoted by $\left\langle...\right\rangle_{\rm ev}$, is taken over events within a given multiplicity class, while $\left\langle\!\left\langle p_{\rm T}\right\rangle\!\right\rangle$ denotes the mean value of $\left\langle p_{\rm T}\right\rangle$ within the same multiplicity class.
    The three-particle correlator, associated with the skewness of the $\left\langle p_{\rm T}\right\rangle$ distribution is expressed as
	\begin{equation}
	\left\langle \Delta p_{\rm T i}\Delta p_{\rm T j}\Delta p_{\rm T k}\right\rangle=\left\langle\frac{\sum_{i, j \ne i, k \ne i, j}^{N_{\rm ch}}(p_{\rm T i}-\left\langle\!\left\langle p_{\rm T}\right\rangle\!\right\rangle ) (p_{\rm T j}-\left\langle\!\left\langle p_{\rm T}\right\rangle\!\right\rangle ) (p_{\rm T k}-\left\langle\!\left\langle p_{\rm T}\right\rangle\!\right\rangle )}{N_{\rm ch}(N_{\rm ch}-1)(N_{\rm ch}-2)} \right\rangle_{\rm ev}		.
	\end{equation}
    \end{widetext}
    However, for practical implementation, these correlators can be written in a more simpler form when expressed in terms of raw moments as follows \cite{skewness} :
	\begin{flalign*}
	\nonumber \langle\!\langle p_{\rm T}\rangle\!\rangle&=\left\langle\frac{Q_1}{N_{\rm ch}}\right\rangle_{\rm ev},
    \end{flalign*}
    \begin{flalign*}
    \nonumber \left\langle \Delta p_{\rm T i} \Delta p_{\rm T j}\right\rangle&=\left\langle \frac{Q_{1}^2-Q_2}{N_{\rm ch}\left(N_{\rm ch}-1\right)}  \right\rangle_{\rm ev}
	-\left\langle\frac{Q_1}{N_{\rm ch}}\right\rangle_{\rm ev}^2, 
    \end{flalign*}
    \begin{flalign*}
     \left\langle \Delta p_{\rm T i} \Delta p_{\rm T j} \Delta p_{\rm T k}\right\rangle =\left\langle \frac{Q_1^3-3Q_2Q_1+2 Q_3}{N_{\rm ch}\left(N_{\rm ch}-1\right)\left(N_{\rm ch}-2\right)} \right\rangle_{\rm ev}\\-3\left\langle \frac{Q_1^2-Q_2}{N_{\rm ch}\left(N_{\rm ch}-1\right)}  \right\rangle_{\rm ev}\left\langle\frac{Q_1}{N_{\rm ch}}\right\rangle_{\rm ev}+2\left\langle\frac{Q_1}{N_{\rm ch}}\right\rangle_{\rm ev}^3,
    \end{flalign*}
	
	Here,  $Q_{\rm n}$s are the moments of the transverse momentum distributions in an event and is written as
	\begin{equation}
	\label{defqn}
	Q_{\rm n}=\sum_{i=1}^{N_{\rm ch}} (p_{\rm T i})^{n}
	\end{equation}
	$p_{\rm T i}$ denotes the transverse momentum of the particle $i$, and the sum runs over all the charged particles in an event. The different value of $n$ corresponds to different orders of the moments. Furthermore, in order to make the analysis less sensitive to the range in $p_{\rm T i}$, the two-particle correlator was normalised with $\left\langle \!\left\langle p_{\rm T}\right\rangle \! \right\rangle$. Thus, the measure of the fluctuations are given by the normalised two-particle correlator,
	$\sqrt{\left\langle \Delta p_{\rm T i}\Delta p_{\rm T j}\right\rangle}/{\left\langle\! \left\langle p_{\rm T}\right\rangle \! \right\rangle}$.

	Likewise, the three-particle correlator is used to give a measure of the skewness as
	\begin{align}
	\gamma_{\left\langle p_{\rm T}\right\rangle}=\frac{\left\langle \Delta p_{\rm T i}\Delta p_{\rm T j} \Delta p_{\rm T k}\right\rangle}{{\left\langle \Delta p_{\rm T i}\Delta p_{\rm T j}\right\rangle}^{3/2}}
	\label{standskew_eq}
	\end{align}
	$\gamma_{\langle p_{\rm T}\rangle}$ is known as standardized skewness and has a system size or centrality dependence which is measured by the number of participant nucleons in the heavy-ion collision process. In an effort to eliminate the trivial size dependence of the skewness, a second measure of the skewness was proposed in the reference \cite{skewness}.
	\begin{align}
	\Gamma_{\left\langle p_{\rm T}\right\rangle}=\frac{\left\langle \Delta p_{\rm T i}\Delta p_{\rm T j} \Delta p_{\rm T k}\right\rangle\left\langle \! \left\langle p_{\rm T}\right\rangle \!\right\rangle}{{\left\langle \Delta p_{\rm T i}\Delta p_{\rm T j}\right\rangle }^2}
	\label{intenskew_eq}
	\end{align}
    $\Gamma_{\langle p_{\rm T}\rangle}$ is called intensive skewness and is independent of the  number of participant nucleons or centrality.
        
    \section{Results and Discussion} \label{result_discussions}
    The analysis utilized 500 million events produced for pp collisions using the PYTHIA8 and HERWIG7 event generators at $\sqrt{s}$=13 TeV. The charged particles were accepted in a pseudorapidity window of $|\eta| \leq 0.8$ with the transverse momentum range $0.15\leq p_{\rm T}\leq 2.0 $ GeV/\textit{c}. The event multiplicity estimation was performed by classifying events based on the charged particle multiplicity obtained in the pseudorapidity range $|\eta|\geq 1.5$. The obtained multiplicity values in the pseudorapidity range $|\eta|\geq 1.5$ scales linearly with the multiplicity of charged particles in the accepted range ( $|\eta| \leq 0.8$) used for the analysis. For each multiplicity class in $|\eta|\geq 1.5$ region, the corresponding mean charged multiplicity $\langle N_{\rm ch} \rangle$ for $|\eta| \leq 0.8$ was obtained to avoid autocorrelation biases.
        \begin{figure*}[!htp]
		\centering
		\includegraphics[width=1.0\linewidth]{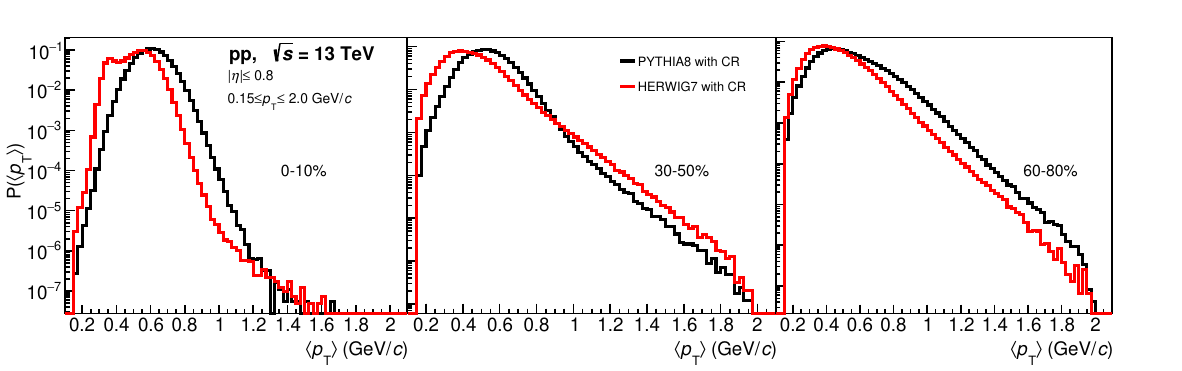}

		\caption{The mean transverse momentum ($\langle p_{\rm T}\rangle$) distributions in different multiplicity classes of events for the PYTHIA8 and HERWIG7 event generators.}
		\label{meanpt_multclass}
     \end{figure*}

 \begin{figure*}[!htp]
		\centering
		\includegraphics[width=0.7\linewidth]{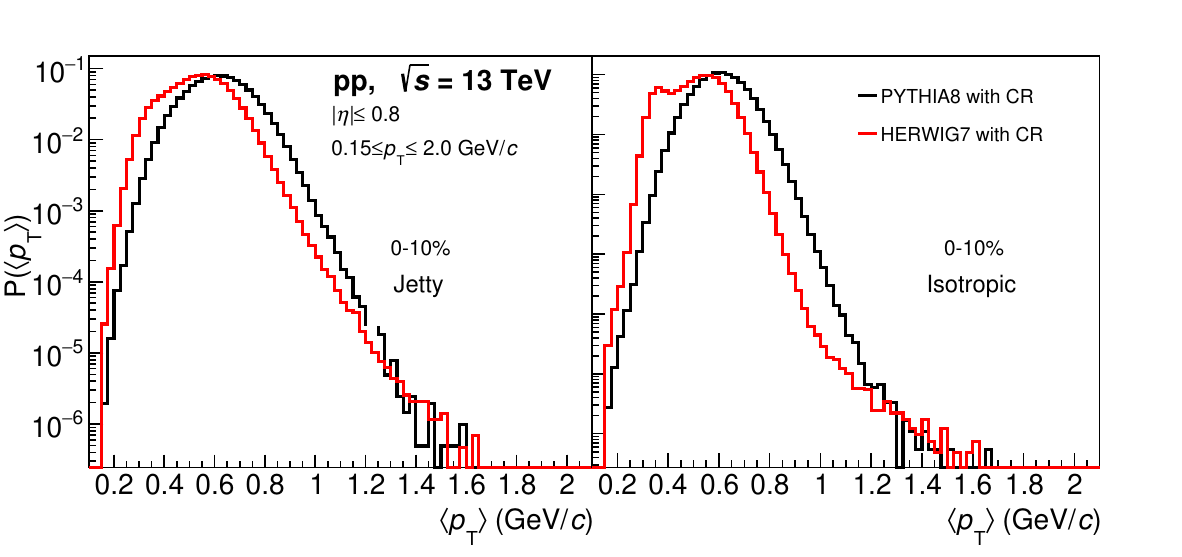}

		\caption{The mean transverse momentum ($\langle p_{\rm T}\rangle$) distributions in jetty and isotropic events for 0-10\% multiplicity class.}
		\label{meanpt_10sphero}
\end{figure*}
    The event-wise $\langle p_{\rm T}\rangle$ distributions for three different multiplicity classes of events: 0-10\%, 30-50\% and 60-80\% are shown in Fig. \ref{meanpt_multclass}. It can be observed that the distributions are asymmetric for both the models indicating a significant departure from the expectations of central limit theorem (gaussianity). The positive skew is more pronounced for low multiplicity (60-80\%) compared to high multiplicity (0-10\%) collisions. Also, the average value of the $\langle p_{\rm T}\rangle$ distribution increases while the width of the distribution decreases when transitioning from low multiplicity (60-80\%) to high multiplicity (0-10\%) events. This essentially indicates that the fluctuations are comparatively less for high multiplicity events. Although the two models predict comparable fluctuations for various multiplicity classes in the chosen range of $p_{\rm{T}}$, it is interesting to note the appearance of an extra tiny peak for the high multiplicity (0-10\%) events at around $\langle p_{\rm T}\rangle$=0.35 GeV/\textit{c} for HERWIG7 model. This observation has been further investigated by disentangling the events with dominating contributions from hard scattering (jetty events) and additional softer processes (isotropic) in high multiplicity (0-10\%) events. Fig. \ref{meanpt_10sphero} depicts the $\langle p_{\rm T}\rangle$ distributions in the jetty and isotropic events for 0-10\% multiplicity class.
    The peak observed in the HERWIG7 model appears in the isotropic event class but is absent in jetty events. This observation can be attributed to the interplay of soft underlying event activities, which are modeled using the multiparton interactions (MPI) and other soft processes in HERWIG7. These observed fluctuations in the $\langle p_{\rm T}\rangle$ distributions are believed to have a dynamical origin and are not merely the result of the event-by-event statistical fluctuations of average multiplicity of charged particles, $ \langle N_{\rm ch}\rangle$. Therefore, the higher-order moments of $\langle p_{\rm T}\rangle$ fluctuations can reveal non-trivial information on the fluctuations \cite{skewness, acharya2024skewness}.

 \begin{figure*}[!htp]
		\centering
		\includegraphics[width=0.9\linewidth]{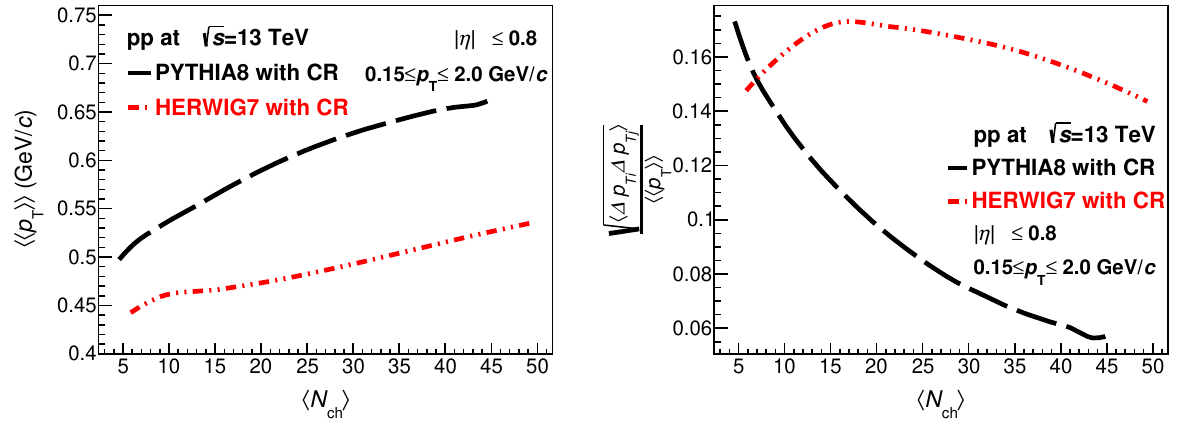}
		\label{mpt_twopart}
		\caption{ The variation of mean transverse momentum $\left\langle\!\left\langle p_{\rm T}\right \rangle\!\right\rangle$ (left) and corresponding fluctuations measurement, $\sqrt{\left\langle \Delta p_{\rm T i}\Delta p_{\rm T j}\right\rangle}/{\left\langle\! \left\langle p_{\rm T}\right\rangle \! \right\rangle}$ (right), in pp collisions at $\sqrt{s}$=13 TeV  plotted as a function of $ \langle N_{\rm ch} \rangle$ for PYTHIA8 and HERWIG7. The kinematic window of acceptance for the charged particles are $|\eta|\leq 0.8$ and $0.2\leq p_{\rm T}\leq 2.0$ GeV/\textit{c}. }
		\label{MB_mpT_twopart}
	\end{figure*}
    
    Fig. \ref{MB_mpT_twopart} shows the predictions for  $\langle\! \langle p_{\rm T} \rangle\!\rangle$ (left), and normalised two-particle correlator, $\sqrt{\left\langle \Delta p_{\rm T i}\Delta p_{\rm T j}\right\rangle}/{\left\langle\! \left\langle p_{\rm T}\right\rangle \! \right\rangle}$ (right), as a function of mean charged particle multiplicity, $\langle N_{\rm ch}\rangle$. As mentioned in Sec. \ref{model_descriptions}, the inclusion of CR processes in both the models directly leads to the increasing trend of $\langle\! \langle p_{\rm T} \rangle\!\rangle$ as a function of average charged particle multiplicity. The mechanism of CR allows the partonic interactions through color strings from different semi-hard scatterings. There is a prior fusion of strings from different multi-partonic systems before hadronization, which leads to a decrease in overall multiplicity, but the particles are more energetic and consequently have a higher value of $\langle \! \langle p_{\rm T} \rangle\!\rangle$. The behavior is qualitatively similar to the one measured by ALICE experiment \cite{mptalice}. Furthermore, the comparison reveals that the $\langle\! \langle p_{\rm T} \rangle\!\rangle$ values as a function of $\langle N_{\rm ch}\rangle$ in PYTHIA8 noticeably deviates from the HERWIG7 model. PYTHIA8 produces significantly higher values of $\langle\! \langle p_{\rm T} \rangle\!\rangle$ for similar multiplicities. The corresponding normalised two-particle transverse momentum correlators, which measure the relative dynamical fluctuations were also compared for the models as a function of average charged particle multiplicity (Fig. \ref{MB_mpT_twopart}). It is evident that both the models produce finite fluctuations values but their characteristic trends deviate significantly as a function of average charged-particle multiplicity. $\sqrt{\left\langle \Delta p_{\rm T i}\Delta p_{\rm T j}\right\rangle}/{\left\langle\! \left\langle p_{\rm T}\right\rangle \! \right\rangle}$ follows an approximate power-law dependence with a reduction of one order of magnitude with increasing multiplicity for PYTHIA8. However, HERWIG7 yields maximum fluctuations around $\langle N_{\rm ch}\rangle \approx$ 15, which diminishes approximately by 3\% for higher multiplicity events. This observation highlights the effect of different hadronization mechanisms on the two-particle correlator.
    \begin{figure*}[!htp]
		\centering
		\includegraphics[width=0.9\linewidth]{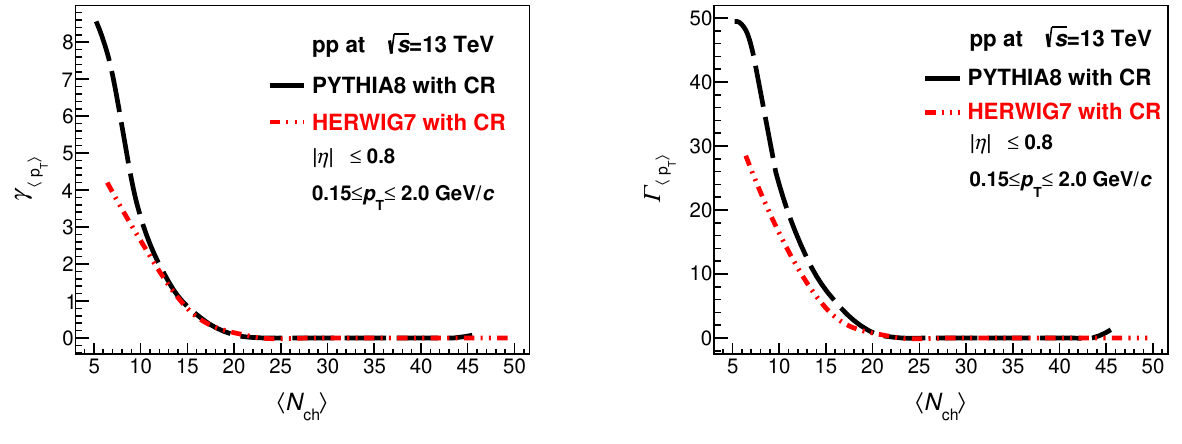}
		\caption{ The different measurements of the skewness shown as a function of $\left\langle N_{\rm ch}\right\rangle$ in pp collisions at $\sqrt{s}$=13 TeV for PYTHIA8 and HERWIG7 models.}
		\label{MB_plot_gammas}
	\end{figure*}
     
    The skewness of the mean transverse momentum distribution is proposed to reflect crucial information on the hydrodynamic expansion of the system created in relativistic nuclear collisions \cite{skewness}. It was suggested that the skewness of $\langle p_{\rm T} \rangle$ results from the fluctuations in the initial energy density of the fluid \cite{gardim2021effects, giacalone2021correlation}. A positive value, larger than the independent baseline for intensive skewness, has been measured by the ALICE experiment in Pb-Pb collisions at $\sqrt{s_{\rm NN}} = 5.02$ TeV \cite{acharya2024skewness}, which is consistent with previous predictions using hydrodynamic model calculations \cite{skewness}. Since a number of physical processes like jets and resonances significantly influence the observed fluctuations in $\langle p_{\rm T} \rangle$, similar measurements are also motivated in pp collisions for an effective understanding of the interplay of various physical processes at the baseline level. Fig. \ref{MB_plot_gammas} shows the model predictions of the standardized and intensive skewness as a function of $\left\langle N_{\rm ch}\right\rangle$ in pp collisions at $\sqrt{s}$ = 13 TeV using PYTHIA8 and HERWIG7 generators. The standardized skewness was anticipated to depend on the number of participant nucleons in Pb$-$Pb collisions, leading to its enhancement in the most central collisions \cite{acharya2024skewness}. However, in pp collisions, both versions of the skewness show a qualitatively similar trend with increasing $\left\langle N_{\rm ch}\right\rangle$. This could likely arise due to the lack of system size fluctuations in pp collisions, in contrast to heavy-ion collisions where the fluctuations in participant nucleons introduce an additional source, thereby making standardized skewness more centrality dependent. Fig. \ref{MB_plot_gammas} shows the estimations of the standardized (left) and intensive (right) skewness of the $\langle \! \langle p_{\rm T} \rangle\!\rangle$ distributions as a function of average multiplicity. Both the models produce comparable skewness except a small deviation towards the low multiplicity events where an asymmetry around the $\langle \! \langle p_{\rm T} \rangle\!\rangle$ distribution is expected due a dominance of hard scatterings. However, observed skewness vanishes beyond  $\langle N_{\rm ch } \rangle \approx 20$ for both the models. This could suggest a model-independent limiting value of multiplicity where the impact of hard scatterings diminishes and the soft QCD-dominated processes like multiparton interactions stabilizes for pp collisions, thereby making the event-by-event fluctuations more Gaussian-like.
    \begin{figure*}[!htp]
		\centering
		\includegraphics[width=1.0\linewidth]{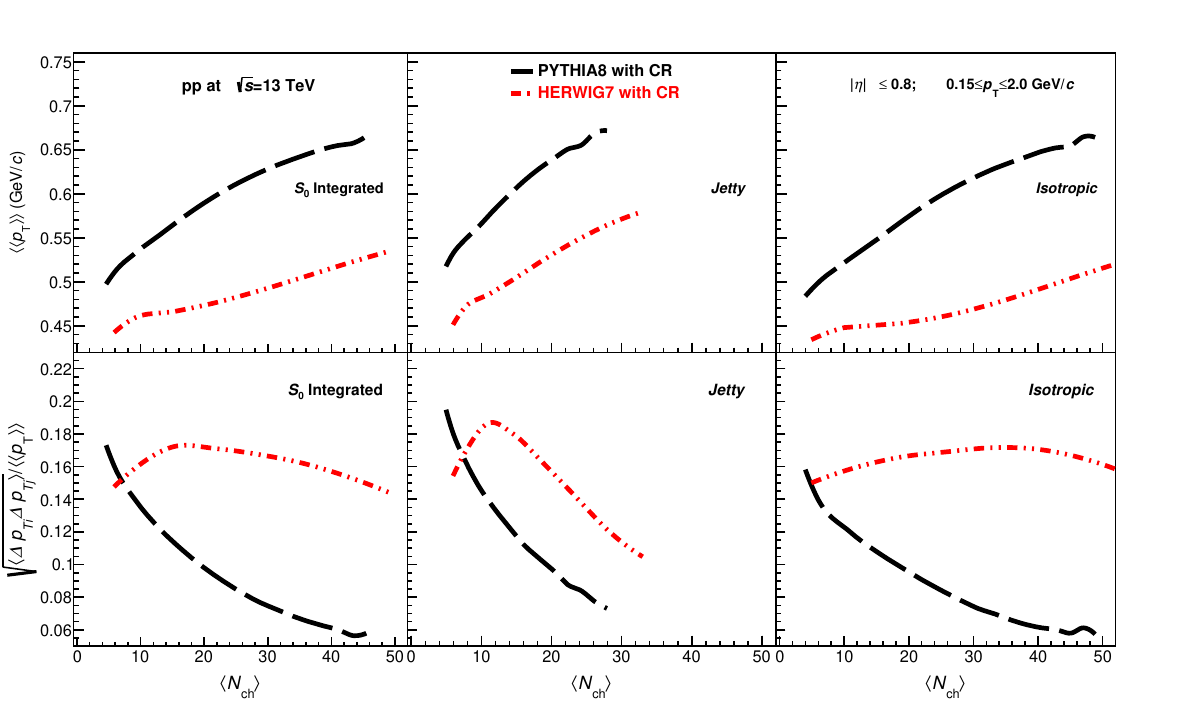}
		\caption{ Comparison of $\left\langle\!\left\langle p_{\rm T}\right \rangle\!\right\rangle$ and normalised two particle correlator, $\sqrt{\left\langle \Delta p_{\rm T i}\Delta p_{\rm T j}\right\rangle}/{\left\langle\! \left\langle p_{\rm T}\right\rangle \! \right\rangle}$ in spherocity integrated (left), jetty (middle) and isotropic (right) events for PYTHIA8 and HERWIG7.}
		\label{sphero_mpT_twopart}
    \end{figure*}

    \begin{figure*}[!htp]
		\centering
		\includegraphics[width=1.0\linewidth]{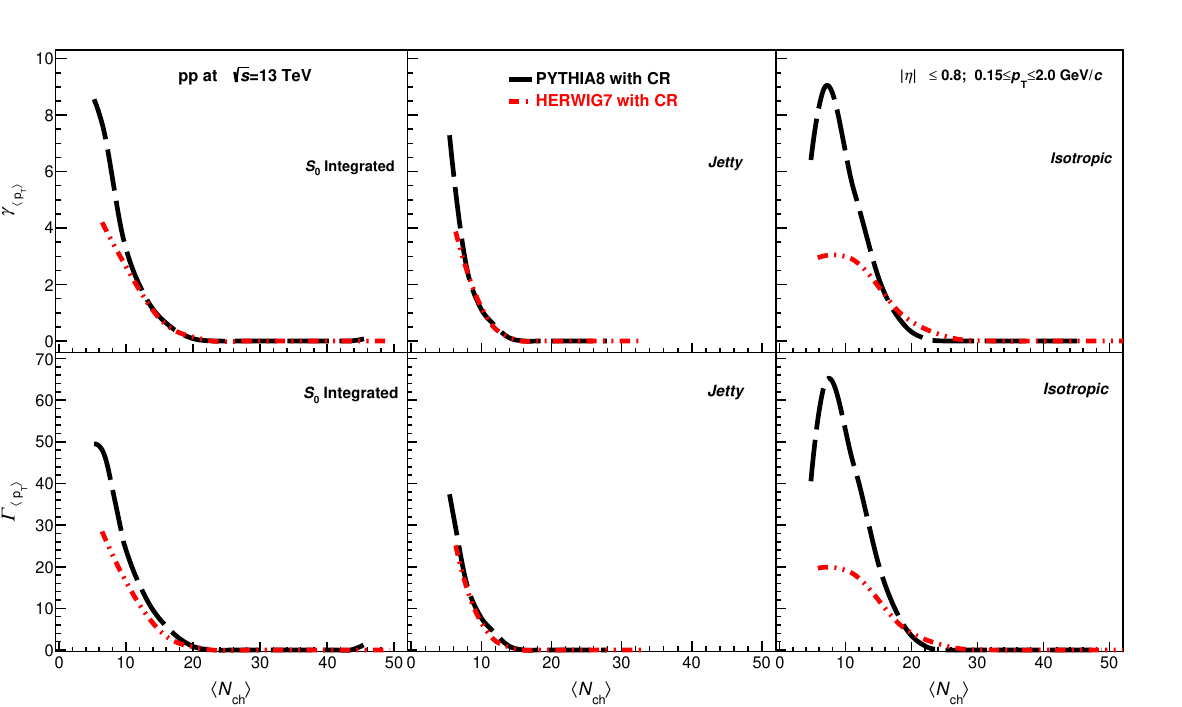}
		\caption{ Different measurements of skewness compared for spherocity integrated (left), jetty (middle) and isotropic (right) events in PYTHIA8 and HERWIG7 models for pp collisons at $\sqrt{s}=13$ TeV.}
		\label{sphero_plot_gammas}
    \end{figure*}

    Comprehending the complex interplay of the various physical processes and their respective contributions to the fluctuation dynamics that affect the particle productions in high energy collisions, is the key component of studying the event-by-event fluctuations. As mentioned in Sec. \ref{intro}, where number of processes dominate the fluctuation dynamics in heavy-ion collision systems, in small systems like pp, the fluctuations are largely driven by jets and various underlying soft processes. Therefore, an analysis of the impact of the jets and associated underlying events (UE) in such collisions becomes essential to effectively describe the observed fluctuations in heavy-ion collisions. This is done by studying the relative strengths of various higher-order moments of the $\langle p_{\rm T} \rangle$ distribution in events with pronounced to less pronounced jets. The transverse spherocity, described in Sec. \ref{spherocity} is used to distinguish events based on their geometric shape in the transverse plane. Events with $S_{0} \rightarrow 0$ feature events with prominent jets, whereas those with $S_{0} \rightarrow 1$ feature events with maximum contribution from soft underlying events. Fig. \ref{sphero_mpT_twopart} presents the comparison of the $\langle\! \langle p_{\rm T} \rangle\!\rangle$ (top) and normalised two-particle  correlator, $\sqrt{\left\langle \Delta p_{\rm T i}\Delta p_{\rm T j}\right\rangle}/{\left\langle\! \left\langle p_{\rm T}\right\rangle \! \right\rangle}$ (bottom), as a function of $\langle N_{\rm ch}\rangle$ for the selected spherocity classes in pp collisions at $\sqrt{s}$=13 TeV for PYTHIA8 and HERWIG7. The left panel shows the $\langle\! \langle p_{\rm T} \rangle\!\rangle$ (top), and normalised two-particle  correlator, $\sqrt{\left\langle \Delta p_{\rm T i}\Delta p_{\rm T j}\right\rangle}/{\left\langle\! \left\langle p_{\rm T}\right\rangle \! \right\rangle}$ (bottom) for spherocity integrated events, i.e. events without spherocity selection; the middle and right panel presents the observables for jetty and isotropic events, respectively. The $S_{0}$ values for the selection of jetty an isotropic events are listed in Table \ref{spherocity_limits}. The presence of jets is found to produce a stronger upward-rising trend in $\langle\! \langle p_{\rm T} \rangle\!\rangle$ towards the high multiplicity events for both the models. Consequently, the strength of the  correlator, $\sqrt{\left\langle \Delta p_{\rm T i}\Delta p_{\rm T j}\right\rangle}/{\left\langle\! \left\langle p_{\rm T}\right\rangle \! \right\rangle}$, shows an enhancement by 20\% for the jet-enriched samples. The increase in correlation strength most likely arises from the fact that jet particles are often more correlated than other particles due to their fragmentation into narrow cone emission. It would be interesting to investigate if this change of 20\% in the correlator strength persists for heavy-ion systems, where a suppression of jets are expected from mid-central to central collisions. This should subsequently lessen the correlation strength in comparison to the scaled reliance brought on by the correlator's dilution in mid-to-central heavy-ion collisions. However, the contributions of numerous mini-jets can also influence the strength as well as the slope of the correlator. Hence, one would be able to infer the possible contributions of jets on the fluctuation observed in heavy-ion collisions. The effects of spherocity selections on the estimated skewness of the  $\langle\! \langle p_{\rm T} \rangle\!\rangle$ distribution are shown in Fig. \ref{sphero_plot_gammas}. The selection of soft UE dominated events (isotropic) result in a reduction on the skewness for HERWIG7. However, the MPI dominated isotropic events enhances the observed skewness for PYTHIA8. \par 
     
     Furthermore, the observed fluctuations exhibit similar slopes for jetty events in both PYTHIA8 and HERWIG7, suggesting that the dominant physics governing these events namely, the hard QCD scattering process is consistently captured by both models. Since jetty events primarily select hard interactions with minimal contributions from the underlying event (UE), the agreement indicates that the differences in parton shower and hadronization models between PYTHIA8 and HERWIG7 do not significantly impact the fluctuation trends in this regime. However, for isotropic events, notable deviations emerge between the two models, pointing to the differences in handling of UE in these models. Since isotropic events are more sensitive to soft QCD effects and multiparton interactions, the observed discrepancies are likely to stem from variations in UE modeling, particularly in the treatment of multiple parton interactions and beam remnants.
    
    \section{Summary} \label{summary}
    Event-by-event fluctuation of the mean transverse momentum ($\langle p_{\rm T} \rangle$) has been studied as a function of the average charged particle multiplicity ($\langle N_{\rm ch} \rangle$) in pp collisions at $\sqrt{s}=13$ TeV using the Monte Carlo generators PYTHIA8 and HERWIG7. The charged particles were considered in the kinematic range $0.15\leq p_{\rm T}\leq 2.0$ GeV/\textit{c} and $|\eta|\leq 0.8$. The finite positive value of the normalised two-particle correlator $\sqrt{\left\langle \Delta p_{\rm T i}\Delta p_{\rm T j}\right\rangle}/{\left\langle\! \left\langle p_{\rm T}\right\rangle \! \right\rangle}$ as a function of the average charged particle multiplicity ($\langle N_{\rm ch} \rangle$) is an indicative of the presence of dynamical fluctuations. The characteristic decrease of the correlator is attributed to independent superposition of different multiparton interactions in pp collisions. However, the estimation of the correlator $\sqrt{\left\langle \Delta p_{\rm T i}\Delta p_{\rm T j}\right\rangle}/{\left\langle\! \left\langle p_{\rm T}\right\rangle \! \right\rangle}$ reports distinct dependencies on the underlying mechanism of hadron formation in the final state. The HERWIG7 produces larger fluctuations than that in PYTHIA8 for similar multiplicity events. Further, the event-shape analysis based on the transverse spherocity reveal that the jet dominated events significantly enhance the correlator strengths for both the models. The skewness has additionally been investigated as function of the average charged particle multiplicity ($\langle N_{\rm ch} \rangle$). A finite positive value of the skewness is reported for low multiplicity events, which subsequently vanishes beyond the multiplicity value, $\langle N_{\rm ch} \rangle$=20 for both the models. This could potentially indicate a model independent  limit for the multiplicity in pp collisions where the impact of soft QCD-dominated processes stabilizes, making the event-by-event fluctuations more Gaussian-like. Finally, the investigation with the transverse spherocity suggests that the different hadronization pictures do not impact the fluctuations in events dominated by hard scattering processes. The difference in sampling the various underlying events like multiparton interactions and beam remnants manifest to the deviations on the observed trends for these models.
    
    \section{Acknowledgement}
    The authors would like to thank the Department of Science and Technology (DST) and University Grants Commission (UGC), India  for supporting the present work. Sadhana Dash would like to acknowledge and thank SERB Power fellowship, {\bf SPF/2022/000014} for supporting the present work.
    
    \bibliography{myref}
    
\end{document}